\documentclass[11pt,a4paper]{article}
\usepackage{float}
\usepackage{multirow}
\usepackage[cp1252,utf8]{inputenc}
\usepackage{hyperref}
\usepackage{graphicx}
\usepackage{wrapfig}
\usepackage{amsmath}
\usepackage{amsfonts}
\usepackage{amssymb}
\usepackage{relsize} 
\usepackage[top=1in, bottom=2cm, left=1.5cm, right=1.5cm]{geometry}
\usepackage{authblk}
\usepackage{ulem}
\usepackage{changepage} 

\usepackage{afterpage}

\usepackage{placeins}

\usepackage{wrapfig}
\usepackage{cite} 
\usepackage[nottoc,notlof,notlot]{tocbibind} 

\title{\bf Mutual information, islands in black holes and the Page curve}
\author[a]{\bf Ashis Saha  \thanks{ashisphys18@klyuniv.ac.in}}
\author[b]{\bf Sunandan Gangopadhyay \thanks{sunandan.gangopadhyay@bose.res.in}}
\author[c]{\bf Jyoti Prasad Saha \thanks{jyotiprasadsaha@gmail.com}}

\affil[a,c]{\textit{Department of Physics, University of Kalyani, Kalyani 741235, India}}
\affil[b]{\textit{Department of Theoretical Sciences,
		S.N.~Bose National Centre for Basic Sciences,}
	\textit{JD Block, Sector-III, Salt Lake, Kolkata 700106, India}}

\date{}

\begin{document}
\maketitle
\begin{abstract}
	\noindent The role played by the mutual information of subsystems on the Page curve is explored in this paper. With the total system consisting of the black hole and radiation, together with the inclusion of island, we observe that the vanishing of mutual information between $B_+$ and $B_-$ which in turn means the disconnected phase of the entanglement wedge corresponding to $B_+\cup B_-$, yields a time scale of the order of scrambling time. This results in a time independent expression for the fine grained entropy of Hawking radiation consistent with the correct Page curve. We also find corrections to this entropy and Page time which are logarithmic and inverse power law in form.
\end{abstract}

\noindent The study of information loss paradox has been one of the most fundamental problems from the point of view of quantum theory of gravity \cite{Hawking:1975vcx,Hawking:1976ra}. For an evaporating black hole, it has been shown that the entropy of radiation monotonically increases with respect to the observer's time. However, the process of unitary evolution demands that this entropy should vanish at the end of the evaporation process. The reason for this is the following. Before the collapse of matter, the state of a quantum field on a full Cauchy slice is pure, and it should remain the same after the evaporation of the black hole. Furthermore, the time dependency of the entropy of radiation is portrayed by the \textit{Page curve} \cite{Page:1993wv,Page:2013dx}. The Page curve efficiently resolves the problem of information loss paradox by introducing a time scale known as the \textit{Page time} $t_p$. The information loss paradox in terms of the Page curve can be understood as follows. The fine grained entropy of Hawking radiation is identified by the von Neumann entropy of quantum fields on the region $R$ outside the black hole. Now assuming the state on the full Cauchy slice to be a pure state, the fine grained entropy of radiation $S(R)=S(R^c)$, where $S(R^c)$ can be understood as the fine grained entropy of the black hole subsystem\footnote{We note that if the region $R$ corresponds to the fine grained entropy of radiation, then the region $R^c$ should correspond to the fine grained entropy of the black hole as the total system comprises of the black hole + radiation.}. This in turn means that $S(R)$ should always satisfy the following property, $S(R)\leq S_{BH}$, where $S_{BH}$ is the coarse grained entropy of the black hole subsystem\footnote{In a strict sense, the coarse grained entropy of the black hole subsystem is $S_{BH}+S_{\textrm{VN}}(QFT)$, where $S_{\textrm{VN}}(QFT)$ represents the von Neumann entropy of quantum fields in a small region outside the event horizon. However, this should be very small.}. However, it has been observed that just after the Page time $t_p$, $S(R)>S_{BH}$ which gives rise to the paradox. Recent studies in this direction \cite{Penington:2019npb,Almheiri:2019psf,Almheiri:2019hni,Almheiri:2019yqk,Almheiri:2019qdq,Chen:2019uhq,Hashimoto:2020cas,Anegawa:2020ezn,Matsuo:2020ypv,Akal:2020twv,Almheiri:2020cfm,Hartman:2020swn,Dong:2020uxp,Balasubramanian:2020xqf,Raju:2020smc,Alishahiha:2020qza} has suggested that the entropy of Hawking radiation receives contribution from certain auxiliary regions known as \textit{islands} with its boundaries identified as surfaces said to be  \textit{quantum extremal surfaces} (QES) \cite{Ryu:2006bv,Engelhardt:2014gca}. This means that at Page time, some non-trivial QES appears in the spacetime which cancels out the time-dependency of $S(R)$, leading to a saturated fine grained entropy of Hawking radiation. Including the island contributions, the fine grained entropy of Hawking radiation reads \cite{Almheiri:2019psf,Almheiri:2019hni}
\begin{eqnarray}\label{equation1}
S(R)=\textrm{min}~\mathop{\textrm{ext}}_{\mathcal{\mathrm{I}}}\bigg\{\frac{\textrm{Area}(\partial I)}{4G_N}+S_{\textrm{matter}}(I\cup R)\bigg\}~.	
\end{eqnarray}\\  
The concept of island formation has risen from the application of replica technique in gravitational backgrounds. Following the replica method to compute the von Neumann entropy of matter fields in region $R$, one constructs the partition function as a gravitational path integral on this replicated geometry \cite{Penington:2019kki,Almheiri:2019qdq,Colin-Ellerin:2020mva,Goto:2020wnk}. One of the saddle points of this path integral is known as the replica wormholes (with correct boundary conditions) which connects different copies of the spacetime. These replica wormholes eventually lead to the island formula (eq.\eqref{equation1}).\\
\noindent In this work, we investigate the role played by mutual information of subsystems in the island formulation. In particular we shall obtain precise conditions under which $S(R)$ assumes a time independent form. We start by considering an eternal BTZ black hole given by the following metric \cite{Banados:1992wn,Saha:2018jjb}
\begin{eqnarray}\label{equation2}
ds^2 = -\frac{(r^2-r_{+}^2)}{l^2}dt^2+\frac{l^2dr^2}{(r^2-r_{+}^2)}+r^2d\phi^2
\end{eqnarray}  
where $l$ is the AdS radius and $r_+$ is the event horizon. 
We now follow the standard technique of gluing non-gravitational flat spacetimes (auxiliary thermal baths) on both sides of the two-sided black hole \cite{Almheiri:2013hfa,VanRaamsdonk:2013sza}. This introduces absorbing boundary conditions for the Hawking quanta. Focussing on the matter part, we assume that the whole system (non-gravitational auxillary thermal baths and the BTZ spacetime) is filled with conformal matter of central charge $c$. 
As we know the BTZ black hole is dual to a thermofield doublet state (TFD) in $2d$ conformal field theory (CFT), holographically this set up can be realized as a highly energetic pure state of $2d$ CFT connected to a CFT which is at its global vacuum. This set up enables us to collect Hawking radiation in the region $R$. In Kruskal coordinates, the BTZ geometry reads
\begin{eqnarray}\label{equation3}
ds^2 = -\frac{du~dv}{\Omega^2}+r^2d\phi^2
\end{eqnarray} 
with the conformal factor $\Omega=\left(\frac{r_+}{l}\right)\frac{1}{(r+r_+)}$~. If we do not consider the contribution of the island, then eq.\eqref{equation1} assumes the usual form $S(R)=S_{\textrm{matter}}(R)$, that is, we need to calculate the von Neumann entropy of quantum fields on $R=R_+\cup R_-$.
 
\begin{wrapfigure}[16]{l}{0.55\textwidth}
	\centering
	\includegraphics[width=0.4\textwidth]{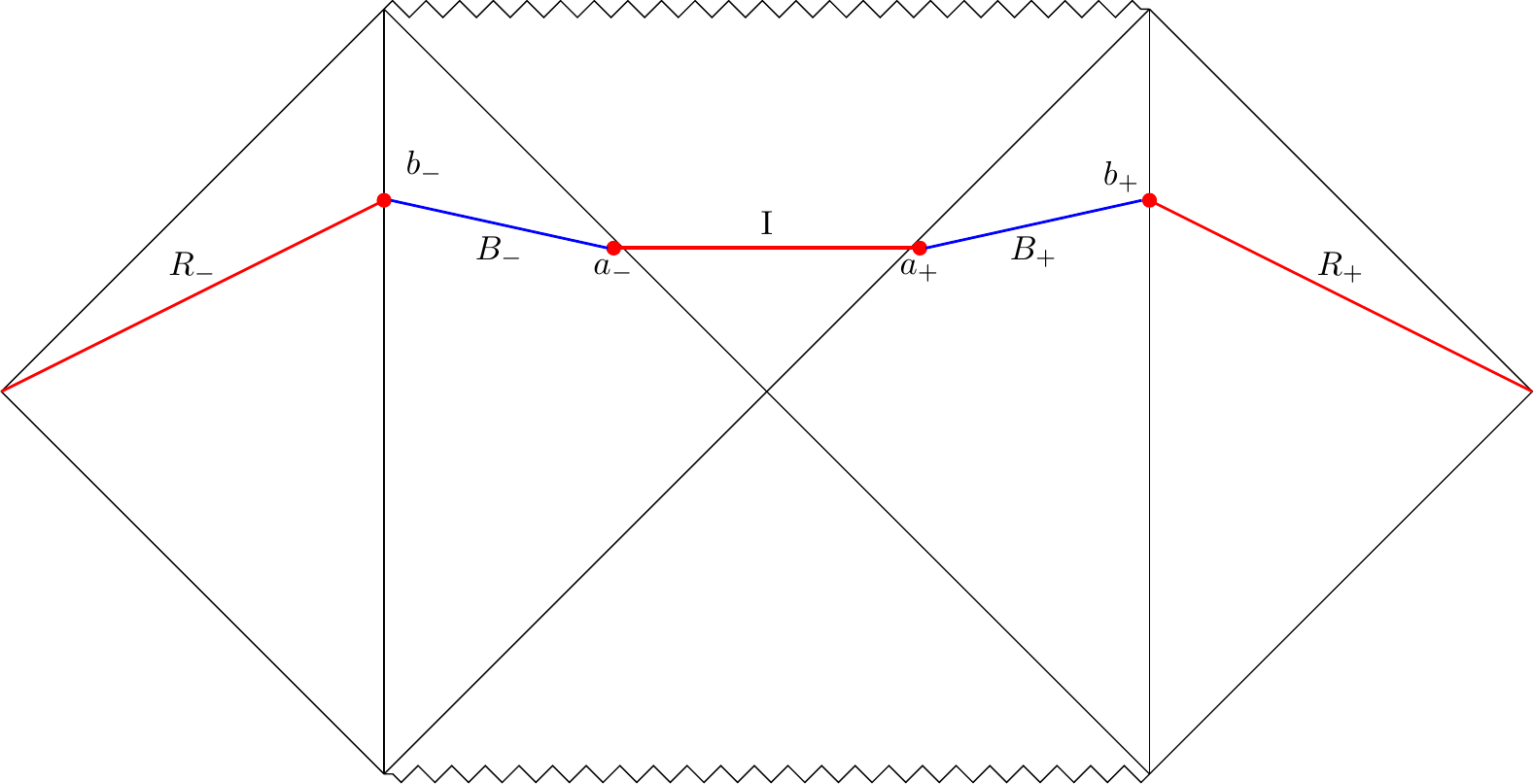}
	\caption{Penrose diagram of eternal BTZ black hole + flat auxiliary thermal bath system specifying regions for the computation of corresponding von Neumann entropies. We denote $a_{\pm}=(\pm t_a,a)$ and $b_{\pm}=(\pm t_b,b)$.}
	\label{fig1}
\end{wrapfigure}		
\noindent Considering the state on the full Cauchy slice to be a pure state, we can write $S_{\textrm{matter}}(R_+\cup R_-) = S_{\textrm{matter}}(B_0)$, where $B_0$ is the region complement to $R=R_+\cup R_-$. 
We now consider only the $s$-wave contribution of the conformal matter \cite{Hashimoto:2020cas,Polchinski:2016hrw,Hubeny:2009rc}, that is, contribution of the massless modes has been considered only in the subsequent analysis. Apart from the massless modes, massive Kaluza-Klein modes are also present. However, the entangling regions in which we are interested are far from each other, this results in a negligible contribution from those massive modes to the von Neumann entropy. Under this approximation, the matter field theory reduces to the effective $2d$ CFT, and hence we use the expression of von Neumann entropy for $2d$ CFT. The finite part of $S(R)$ is given as \cite{Calabrese:2009qy}
\begin{eqnarray}\label{2dCFT}
S_{\textrm{matter}}(R)=\frac{c}{3}\log d(b_+,b_-)
\end{eqnarray}
where the points $b_{\pm}=(\pm t_b,b)$ are shown in Figure \eqref{fig1}.
For the BTZ metric, the distance between the points $b_+$ and $b_-$ is given by
\begin{eqnarray}
d(b_+,b_-)&=&\sqrt{\frac{\left[u(b_-)-u(b_+)\right]\left[v(b_+)-v(b_-)\right]}{\Omega(b_+)\Omega(b_-)}}\nonumber\\
&=&\sqrt{\left(\frac{4l^2}{r_+^2}\right)(b^2-r_+^2)\cosh^2\left(\frac{r_+t_b}{l^2}\right)}~.
\label{dist}
\end{eqnarray}
Substituting this in eq.(\ref{2dCFT}) gives 
\begin{eqnarray}\label{equation4}
S_{\textrm{matter}}(R)=\left(\frac{c}{3}\right)\log\left[\left(\frac{\beta}{\pi l}\right)\sqrt{b^2-r_+^2}\cosh\left(\frac{2\pi t_b}{\beta}\right)\right]~~
\end{eqnarray}
where $c$ is the central charge. It can be observed that at late times ($t_b\gg\beta$), the above expression has a linear time dependency
\begin{eqnarray}\label{lin}
S(R)\equiv S_{\textrm{matter}}(R)\approx \frac{c}{3}\left(\frac{2\pi t_b}{\beta}\right). 
\end{eqnarray}
This in turn means that it will be much greater than the coarse grained entropy of the black hole at late times. As mentioned earlier, in order to resolve this issue, the contribution of the island has to be taken into account. Using the fact that the matter entropy part in eq.\eqref{equation1} satisfies the property $S_{\textrm{matter}}(I\cup R)=S_{\textrm{matter}} (B_+\cup B_-)$ (see Fig.\eqref{fig1}) and using the formula for von Neumann entropy corresponding to two disjoint intervals ($B_+$ and $B_-$), the finite part of $S_{\textrm{matter}}(B_+\cup B_-)$ is given by \cite{Casini:2009sr,Calabrese:2009ez}
\begin{eqnarray}\label{equation5}
S_{\textrm{matter}}(B_+\cup B_-)=\left(\frac{c}{3}\right)\log\Big[\frac{d(a_+,a_-)d(b_+,b_-)d(a_+,b_+)d(a_-,b_-)}{d(a_+,b_-)d(a_-,b_+)}\Big]
\end{eqnarray} 
where $a_{\pm}=(\pm t_a,a)$ is shown in Figure \eqref{fig1}. The expressions for each of the distances in the above expression can be computed from eq.\eqref{equation3} and reads\footnote{The expression corresponding to $d(b_+,b_-)$ is given in eq.\eqref{dist}.}
\begin{eqnarray}
	d(a_+,a_-)&=& \sqrt{\left(\frac{4l^2}{r_+^2}\right)(a^2-r_+^2)\cosh^2\left(\frac{r_+t_a}{l^2}\right)}\nonumber\\
	d(a_+,b_+)=d(a_-,b_-)&=&\sqrt{\frac{(a+r_+)(b+r_+)}{(\frac{r_+}{l})^2}}\bigg[\left(\frac{b-r_+}{b+r_+}\right)+\left(\frac{a-r_+}{a+r_+}\right)-2\sqrt{\left(\frac{a-r_+}{a+r_+}\right)\left(\frac{b-r_+}{b+r_+}\right)}\nonumber\\
	&&\times\cosh\left(\frac{r_+}{l^2}(t_a-t_b)\right)\bigg]^{\frac{1}{2}}\nonumber\\
	d(a_+,b_-)=d(a_-,b_+)&=&\sqrt{\frac{(a+r_+)(b+r_+)}{(\frac{r_+}{l})^2}}\bigg[\left(\frac{b-r_+}{b+r_+}\right)+\left(\frac{a-r_+}{a+r_+}\right)+2\sqrt{\left(\frac{a-r_+}{a+r_+}\right)\left(\frac{b-r_+}{b+r_+}\right)}\nonumber\\
	&&\times\cosh\left(\frac{r_+}{l^2}(t_a+t_b)\right)\bigg]^{\frac{1}{2}}~.
\end{eqnarray}
 Following the recent approach in this direction, at late times ($t_a, t_b \gg \beta$), the above expression of $S_{\textrm{matter}}(B_+\cup B_-)$ takes the following form (neglecting terms $\mathcal{O}(e^{-\frac{2\pi t}{\beta}})$) \cite{Matsuo:2020ypv}
\begin{eqnarray}\label{equation6}
S_{\textrm{matter}}(B_+\cup B_-) \approx S_{\textrm{matter}}(B_+)+ S_{\textrm{matter}} (B_-).
\end{eqnarray} 
Using this, eq.\eqref{equation1} simplifies to
\begin{eqnarray}\label{equation7}
S(R)\approx\textrm{min}~\mathop{\textrm{ext}}_{\mathcal{\mathrm{I}}}\bigg\{\frac{\textrm{Area}(\partial I)}{4G_N}+S_{\textrm{matter}}(B_+)+ S_{\textrm{matter}} (B_-)\bigg\}.~~~~
\end{eqnarray}
Standard extremization of the island parameters ($t_a$ and $a$) yields $t_a=t_b$ and $a\approx r_+$~. Substitution of these extremized values in eq.\eqref{equation7} leads to the result $S(R)\approx2S_{BH}+...$~.\\
\noindent This gives the correct Page curve and therefore resolves the information loss paradox. We shall now explore the role of mutual information of subsystems in the island formula. Before we delve into this, we would like to point out that although the `replica wormhole' saddle points are present, the `Hawking saddle' of the gravitational path integral dominates upto Page time. This leads to a linearly time-dependent profile for $S(R)$. However, just after the Page time, the replica wormhole saddle points start to dominate and in order to compute the $S(R)$ correctly, one needs to consider the island contributions \cite{Almheiri:2019qdq,Colin-Ellerin:2020mva,Goto:2020wnk}. On the other hand, in order to get the expression given in eq.\eqref{equation6}, one needs to ignore all the contributions from the terms with time dependency $\mathcal{O}(e^{-\frac{2\pi t}{\beta}})$ \cite{Hashimoto:2020cas}. This can be noted from the following expression of $S_{\textrm{matter}}(B_+:B_-)$. This reads
\begin{eqnarray}
S_{\textrm{matter}}(B_+:B_-)&=&S_{\textrm{matter}}(B_+)+S_{\textrm{matter}}(B_-)+\left(\frac{c}{3}\right)\log(1+e^{-\frac{4\pi t_a}{\beta}})+\left(\frac{c}{3}\right)\log(1+e^{-\frac{4\pi t_b}{\beta}})\nonumber\\
&-&\left(\frac{c}{3}\right)\log\left(1+e^{-\frac{4\pi (t_a+t_b)}{\beta}}+\sqrt{\frac{b-r_+}{b+r_+}}\sqrt{\frac{a+r_+}{a-r_+}}e^{-\frac{2\pi (t_a+t_b)}{\beta}}\right)~. 
\end{eqnarray}
From the above expression it can be observed that the relation given in eq.\eqref{equation7} is valid only at leading order. This approximation is not a very reliable approach to proceed further in this current scenario as the soul of the problem is regarding the time-dependency and in a strict sense, this will lead to a time-dependent expression of $S(R)$. Keeping this in mind and taking a clue from eq.\eqref{equation6}, we give the following proposal\footnote{Note that vanishing of the mutual information between $B_+$ and $B_-$, that is, $I(B_+:B_-)=0$ 
does not imply the vanishing of the 
mutual information between $R$ and $I$, that is, $I(R:I)=0$, since otherwise using the fact that $S(B_{+}\cup B_{-})=S(R\cup I)$ it would imply that
$S(B_{+})+S(B_{-})=S(R)+S(I)$ which cannot be true as can be seen using the Ryu-Takayanagi holographic entanglement entropy formula.}.\\

\noindent\textit{Just after the Page time, inclusion of the island contribution leads to the exact vanishing of the mutual correlation between $B_+$ and $B_-$, that is, $I(B_+:B_-)=0$.}\\

\noindent We now give a simple holographic interpretation of our proposal. In the context of holography, the mutual information is a crucial parameter which determines the phase of the entanglement wedge. In other words, $I(A:B)\neq 0$ means a connected (phase) entanglement wedge of $A\cup B$ and $I(A:B)=0$ means disconnected (phase) entanglement wedge of $A\cup B$ \cite{Takayanagi:2017knl}. So we can recast our proposal in the language of the entanglement wedge. We propose that just after the Page time (when the replica wormholes are the dominating saddle points), inclusion of the island contribution leads to the disconnected phase of the entanglement wedge corresponding to $B_+ \cup B_-$. In other words, the island separates the entanglement wedge of $B_+ \cup B_-$ domain, just after the Page time. We shall see that this condition for saturation of mutual information $I(B_+:B_-)$ results in a time difference of the order of scrambling time which then gives a time independent expression for $S(R)$. We start by computing the mutual information $I(B_+:B_-)$. This by definition is given by the following expression
\begin{eqnarray}\label{equation8}
I(B_+:B_-) = S_{\textrm{matter}}(B_+) +	S_{\textrm{matter}}(B_-) -	S_{\textrm{matter}}(B_+\cup B_-)
\end{eqnarray}  
where $S_{\textrm{matter}}(B_\pm)=\frac{c}{3} \log d(a_\pm,b_\pm)$. Computing the right hand side of the above expression, we observe that $I(B_+:B_-)$ 
vanishes if the following condition is satisfied
\begin{eqnarray}\label{equation9}
t_a-t_b=|r^*(a)-r^*(b)|
\end{eqnarray} 
where $r^*(r)$ is the ``tortoise coordinate"
\begin{eqnarray}
	r^*(r)&=&\int \frac{dr}{f(r)}\nonumber\\
	&=& \frac{l^2}{2r_+}\left[\int\frac{dr}{(r-r_+)}-\int\frac{dr}{(r+r_+)}\right]\nonumber\\
	&=&\left(\frac{l^2}{2r_+}\right)\log\left(\frac{r-r_+}{r+r_+}\right)~.
\end{eqnarray}
Note that the saturation of the mutual information between $B_+$ and $B_-$ relates the difference in time between $t_a$ and $t_b$. Hence, $t_a$ gets determined in terms of $t_b$ and $|r^*(a)-r^*(b)|$. This is one of the main findings in this paper. This differs from the earlier approach where $t_a$ gets fixed from extremization condition and gives rise to $t_a=t_b$. Substituting $t_a=t_b+|r^*(a)-r^*(b)|$ in $S_{\textrm{matter}}(I\cup R)$, we obtain
\begin{eqnarray}\label{equation10}
S_{\textrm{matter}}(I\cup R) = \frac{c}{3}\log\left[\left(\frac{2l^2}{r_+^2}\right)\sqrt{b^2-r_+^2}\sqrt{a^2-r_+^2}\right].~~~~~
\end{eqnarray}
It is remarkable to observe that the obtained expression of $S_{\textrm{matter}}(I\cup R)$ has no time dependency at all. The mutual information saturation condition has removed the dependency on time completely. This is in contrast to the extremization approach which yields $t_a=t_b$, which in turn yields a time dependent expression for $S(R)$. We now need to find the extremized value of $``a"$. This would fix the positions of the QES. By substituting eq.\eqref{equation10} in eq.\eqref{equation1} and using the fact $\frac{\textrm{Area}(\partial I)}{4G_N}=2\times\frac{2\pi a}{4G_N}$, the extremization condition $\partial_a S(R)=0$ yields
\begin{eqnarray}\label{equation11}
\frac{\pi}{G_N}+\left(\frac{c}{3}\right)\frac{a}{a^2-r_+^2}=0.
\end{eqnarray}
Solving the above equation perturbatively leads us to the following expression for $``a"$
\begin{eqnarray}\label{equation12}
a = r_+ + \left(\frac{cG_N}{6\pi}\right)^2\frac{1}{2r_+}+...~.
\end{eqnarray}
Note that the correction term to $``a"$ suggests that the QES is formed just outside the horizon of the BTZ black hole.
We now substitute the above extremized value of $``a"$ in $S(R)$ and obtain the final expression corresponding to the fine grained entropy of Hawking radiation. This reads
\begin{eqnarray}\label{equation13}
S(R) = 2S_{BH}-\frac{2c}{3}\log\left(S_{BH}\right)+\frac{\left(\frac{c}{6}\right)^2}{4S_{BH}}+\frac{\left(\frac{c}{6}\right)^3}{6S_{BH}^2}+...
\end{eqnarray}
where $S_{BH}$ is given by
\begin{eqnarray}
	S_{BH}=\frac{\pi r_+}{2G_N}~.
\end{eqnarray}
The above expression have some striking features. We observe that apart from the leading piece $2S_{BH}$, the expression contains universal corrections involving the Hawking entropy of the black hole. This comes as a pleasant surprise as these corrections represent signatures of quantum gravity. The fact that the above result gives the correct Page curve and hence solves the information loss paradox has a nice interpretation in terms of mutual information. With the expression of $S(R)$ (given in eq.\eqref{equation13}) in hand and keeping in mind our proposal regarding $I(B_+:B_-)$, we now make the following statement.\\
Just after the Page time, the auxiliary region appearing inside the black hole interior denoted as the island results in a disconnected phase for the entanglement wedge of black hole subsystem $B_+ \cup B_-$. Further, as a consequence of this disconnected entanglement phase of $B_+ \cup B_-$, the time-dependency in $S(R)$ vanishes completely. We now return to the condition given in eq.\eqref{equation9}. Substituting the expression of $``a"$ (given in eq.\eqref{equation12}), we obtain
\begin{eqnarray}\label{equation14}
t_a-t_b &=& \left(\frac{\beta}{2\pi}\right)\log\left(S_{BH}\right)+\left(\frac{\beta}{16\pi}\right)\left(\frac{c}{12}\right)^2\frac{1}{S_{BH}^2}+...\nonumber\\
&\approx& t_{scr}+\left(\frac{\beta}{16\pi}\right)\left(\frac{c}{12}\right)^2\frac{1}{S_{BH}^2}+...~.
\end{eqnarray}
We now make some comments on the above finding. First of all, we note that the leading piece of the above equation is found to be the \textit{scrambling time} ($t_{scr}$) \cite{Sekino:2008he,Hayden:2007cs} and the sub-leading corrections are of the order $\mathcal{O}\left(1/S_{BH}^2\right)$. This in turn means that as soon as the time difference $t_a-t_b$ equals the scrambling time $t_{scr}$, the mutual correlation between $B_+$ and $B_-$ vanishes which results in a time independent nature of $S(R)$. Previous studies stressed that $t_a-t_b$ is very small and hence can be neglected. We have shown that this time difference is precisely the scrambling time and it arises as a consequence of the vanishing mutual information between the black hole subsystems $B_+$ and $B_-$. The scrambling time can also be understood in a physical way. 
If a message is sent from the point $r=b$ towards the island, the time it takes
to reach the island at the earliest is the scrambling time $t_{scr}=t_b -t_a$. Hence,
it is clear that the condition $t_b =t_a$ is an approximate result since $t_b$ should not be equal to $t_a$ in general as it takes a finite amount of time for the information to reach the island from $r=b$. We observe that the exact vanishing of the mutual information between $B_{+}$ and $B_{-}$ actually leads to the non-trivial result  $t_{b}-t_{a}=t_{scr}$. \\

\noindent We now provide a simple generalization of this condition based upon the subadditivity property of von Neumann entropy. The subadditivity property states that if $A$ and $B$ are two subsystems, then the corresponding von Neumann entropies satisfy the property $S(A)+S(B)-S(A\cup B)\geq0$. This in turn represents the non-negative nature of mutual information $I(A:B)\geq0$. By using this property in the present scenario, we can write down the following condition
\begin{eqnarray}\label{equation15}
|r^*(a)-r^*(b)|\geq t_a-t_b~.
\end{eqnarray}
This means that as long as the spatial differences of $B_+$ and $B_-$ are large than their time difference (that is the island is very small and well inside the event horizon), there exists a finite mutual correlation between $B_+$ and $B_-$. This in turn means that the resulting expressions of $``a"$ and $S(R)$ will be time dependent as the entanglement wedge corresponding to $B_+ \cup B_-$ is in connected phase. With time evolution, the island contribution grows and hence $t_a-t_b$ increases and $|r^*(a)-r^*(b)|$ decreases. The island stops growing when the equality is satisfied. The time difference $t_a-t_b$ can be thought as the elapsed for a particle to reach $r^*(a)$ from $r^*(b)$. We now compute the expression for Page time $t_p$. As we know at $t_b=t_p$ (where $t_p\gg \beta$), expressions of $S(R)$ (given in eq.\eqref{lin} and eq.\eqref{equation13}) has to be equal. This leads to the following
\begin{eqnarray}\label{equation16}
t_p = \left(\frac{3\beta}{\pi c}\right) S_{BH}- \left(\frac{\beta}{\pi}\right)\log\left(S_{BH}\right)+\left(\frac{c}{12}\right) \frac{\beta}{8\pi S_{BH}}+...~~~~
\end{eqnarray} 
The leading term of the above expression is the familiar Page time $t_p$. The rest of the terms are sub-leading corrections. 

\noindent We now proceed to discuss another well known system. This is the asymptotically flat eternal Schwarzschild black hole in $3+1$ dimensions. By following the same approach, we note that the condition for vanishing mutual information ($I(B_+:B_-)=0$) is the same which we have given in eq.\eqref{equation9}. Furthermore, the extremization condition leads to the following expression for $``a"$
\begin{eqnarray}\label{equation17}
a=r_+ - \left(\frac{cG_N}{24\pi}\right)\frac{1}{r_+}-\left(\frac{1}{32}\right)\left(\frac{cG_N}{3\pi}\right)^2\frac{1}{r_+^3}+...
\end{eqnarray} 
where we have kept terms upto $\sim\mathcal{O}\left(cG_N\right)^2$. The fine grained entropy of radiation (with the extremized value of $``a"$) is obtained to be
\begin{eqnarray}\label{equation18}
S(R) = 2S_{BH}-  \left(\frac{c}{6}\right)\log(S_{BH})+\frac{\left(\frac{c}{6}\right)^2}{2S_{BH}^2}+...~.
\end{eqnarray}
Similar to the BTZ case, the above expression contains the universal corrections of black hole entropy along with the leading piece $2S_{BH}$. The extremized values of $``a"$ in the above case suggests that the island end points reside slightly inside the event horizon, whereas for the BTZ black hole the end points of the island, that is, the position of the QES lies slightly outside the event horizon. With the value of $``a"$, the condition $I(B_+:B_-)=0$ takes the form
\begin{eqnarray}\label{equation19}
t_a - t_b \approx t_{scr}- \left(\frac{c}{12}\right) \frac{\frac{\beta}{4\pi}}{S_{BH}}+...~.
\end{eqnarray}
Once again we get the scrambling time ($t_{scr}$) as the leading term. The expression for the Page time in this case is found to be
\begin{eqnarray}\label{equation20}
t_p = \left(\frac{3\beta}{\pi c}\right)S_{BH}- \left(\frac{\beta}{4\pi}\right)\log\left(S_{BH}\right)+\left(\frac{c}{12}\right) \frac{\beta}{4\pi S_{BH}}+...~.
\end{eqnarray}
The above expression has a similar form with the one obtained for BTZ black hole.

\noindent We now summarize our findings. In this paper, we show that mutual information plays a very important role in finding the correct Page curve for the fine grained entropy of Hawking radiation. From previous studies in this directon which involved the computation of the correct Page curve for a black hole glued to a pair of non-gravitating thermal auxiliary baths, we observe that the late time approximation (given in eq.\eqref{equation6}) implies that the mutual information $I(B_+:B_-)$ vanishes in the leading order and this leads to extremization condition $t_a\approx t_b$. In this work we have exploited this observation to show that the mutual information $I(B_+:B_-)$ should vanish exactly and the associated extremization condition leads to $t_a-t_b=t_{scr}+...$. The reason for proposing the exact saturation of $I(B_+:B_-)$ is the following.\\
 We observe that upto the Page time, the Hawking saddle dominates and yields a time-dependent profile for $S(R)$. However, just after the Page time, the replica wormholes dominate and the emergence of the island in the black hole interior yields a disconnected phase for the entanglement wedge corresponding to $B_+ \cup B_-$ and as a consequence we propose that the mutual correlation between $B_+$ and $B_-$ should vanish (not only in leading order as shown in previous studies but in all order). A similar observation for the doubly holographic set up can be found in \cite{Grimaldi:2022suv}. Although the doubly holographic framework cannot be applied in this case, we propose that the vanishing of the mutual information between $B_{+}$ and $B_{-}$ would still apply. A possible reason for this is that it leads to the non-trivial result $t_b -t_a =t_{scr}$. Our proposal supports the observation that the mutual information acts as a measure of connectivity of the associated entanglement wedge.\\
 The net result of this vanishing mutual information (or disconnected entanglement wedge) leads to a time-independent expression for $S(R)$, associated with the simple condition $t_a-t_b=|r^*(a)-r^*(b)|$. In addition to this, the position of these surfaces also get determined by the condition of vanishing mutual information. We also present a nice physical interpretation for this condition. Focusing on the condition of vanishing mutual information between two subsystems ($B_+$ and $B_-$) leads to a condition relating the difference in times and difference in positions between the island and region of Hawking radiation collection boundaries. The subadditivity property of von Neumann entropy then implies that as long the spatial distance is greater than the time difference, the mutual correlation between the subsystems $B_+$, $B_-$ is non-zero and the entanglement wedge corresponding to $B_+ \cup B_-$ is in connected phase, this in turn means that the expression of $S(R)$ will be time dependent. Surprisingly we observe that the condition $|r^*(a)-r^*(b)|\geq t_a-t_b$ can be recast to the form $t_a-t_b\leq t_{scr}$ after the extremization of $``a"$ where $t_{scr}$ is the scrambling time. We emphasize that this is a very interesting result as this means that as long as the the time difference $t_a-t_b$ is smaller than the Scrambling time, $S(R)$ will be time dependent, but as soon as $t_a-t_b = t_{scr}$, $S(R)$ will be time independent due to the fact that $I(B_+:B_-)=0$ and the associated entanglement wedge is in disconnected phase. We also retain sub-leading corrections to $``a"$, and study the role played by these sub-leading corrections. By using the extremized value of $``a"$, we observe that $S(R)$ contains universal corrections of Hawking entropy along with the leading piece $2S_{BH}$. This observation is completely new and very interesting as these represent the signatures of quantum gravity. We have also computed the corrections to the Page time $t_p$. We have carried these out for the eternal BTZ black hole (glued to auxiliary thermal baths) and the eternal Schwarzschild black hole.\\
  Although in this paper we have used free CFT but we conjecture that the observation $I(B_+:B_-)=0$ (just after the Page time) should be model independent and consistent as the sole reason behind this is the emergence of the island (or in other words domination of the replica wormholes). Furthermore, the vanishing of mutual information basically defines what the island actually is and what is its role in 
getting the Page curve.\\
  As a future investigation, it will be really exciting to look into the case of evaporating black holes. We believe that the saturation of mutual information of subsystems would also play a central role in these gravitating systems.
  

\section*{Acknowledgements}
AS would like to acknowledge the support by Council of Scientific and Industrial Research (CSIR, Govt. of India) for the Senior Research Fellowship. The authors would like to thank the anonymous referee for very useful comments. 

\bibliographystyle{elsarticle-num}  
\bibliography{References} 


\end{document}